\documentclass[sigconf]{acmart}
\usepackage{enumitem} 
\usepackage{multirow}
\usepackage{soul}  
\usepackage{algorithm}
\usepackage{algorithmic}
\usepackage{amsmath}
\usepackage{booktabs}     
\usepackage{graphicx}     
\usepackage{array}        
\usepackage{caption} 
\usepackage{tcolorbox}

\DeclareMathOperator*{\argmax}{arg\,max}  

\AtBeginDocument{%
  }

\setcopyright{acmlicensed}
\copyrightyear{2025}
\acmYear{2025}
\acmDOI{10.1145/3767695.3769486}
\acmConference[SIGIR-AP 2025]{Make sure to enter the correct
  conference title from your rights confirmation emai}{December 7-10, 2025}{Xi'an, China}

\acmISBN{979-8-4007-2218-9/2025/12.}

\begin{document}

\title{
CogPlanner: Unveiling the Potential of Agentic Multimodal Retrieval Augmented Generation with Planning
}

\author{Xiaohan Yu$^{*}$}
\affiliation{%
  \institution{Huawei Cloud BU}
  \city{Beijing}
  \country{China}}
\email{yuxiaohan5@huawei.com}

\author{Zhihan Yang$^{*}$}
\affiliation{%
  \institution{Huawei Cloud BU}
  \city{Beijing}
  \country{China}}
\email{yangzhihan4@huawei.com}

\author{Chong Chen$^{\dagger}$}
\affiliation{%
  \institution{Huawei Cloud BU}
  \city{Beijing}
  \country{China}}
\email{chenchong55@huawei.com}

\renewcommand{\shortauthors}{Trovato et al.}

\begin{abstract}
Multimodal Retrieval Augmented Generation (MRAG) systems have shown promise in enhancing the generation capabilities of multimodal large language models (MLLMs). However, existing MRAG frameworks primarily adhere to rigid, single-step retrieval strategies that fail to address real-world challenges of information acquisition and query reformulation. In this work, we introduce the task of Multimodal Retrieval Augmented Generation Planning (MRAG Planning) that aims at effective information seeking and integration while minimizing computational overhead. 
Specifically, we propose CogPlanner, an agentic plug-and-play framework inspired by human cognitive processes, which iteratively determines query reformulation and retrieval strategies to generate accurate and contextually relevant responses. CogPlanner supports parallel and sequential modeling paradigms. Furthermore, we introduce CogBench, a new benchmark designed to rigorously evaluate the MRAG Planning task and facilitate lightweight CogPlanner integration with resource-efficient MLLMs, such as Qwen2-VL-7B-Cog. Experimental results demonstrate that CogPlanner significantly outperforms existing MRAG baselines, offering improvements in both accuracy and efficiency with minimal additional computational costs. 
\end{abstract}




\maketitle
\footnotetext[1]{$^{*}$These authors contributed equally to this work.}
\footnotetext[2]{$^{\dagger}$Corresponding author.}

\section{Introduction}

\begin{figure}
    \centering
    \includegraphics[width=1.0\linewidth]{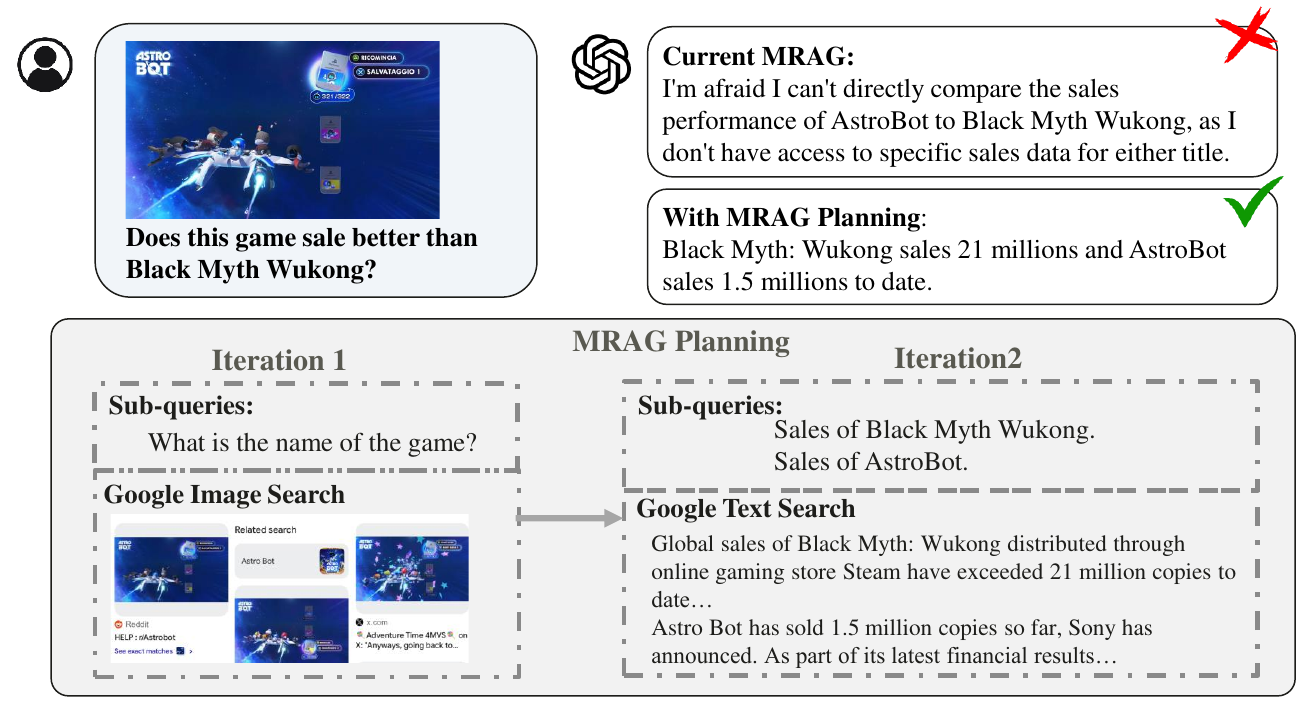}
    \caption{An example of current MRAG system with the benefits of incorporating MRAG Planning.}
    \label{fig:intro}
\end{figure}

Retrieval-Augmented Generation (RAG) has been shown to significantly enhance the performance of large language models (LLMs) by grounding generation in retrieved knowledge \cite{rag1,rag2}. More recently, the emergence of agentic RAG frameworks, exemplified by web agents \cite{li2025websailor,wu2025webdancer}, has highlighted the potential of autonomous reasoning and information seeking capabilities within RAG systems. Howevewr, the increasing demands of real-world applications have necessitated a natural extension of RAG beyond purely texts to encompass multimodal data (e.g., images, videos). This development has led to the advent of Multimodal Retrieval-Augmented Generation (MRAG) \cite{riedler2024beyond}, which equips multimodal large language models (MLLMs) with the ability to retrieve and exploit external multimodal knowledge sources, thereby reducing hallucinations and improving reliability \cite{mrag_survey}.
Existing MRAG frameworks generally adhere to a rigid pipeline characterized by a predetermined retrieval action, either exclusively textual or exclusively visual \cite{chen2022murag,zhu2024murar}, which manifests several critical limitations: 

\begin{itemize}[leftmargin=*]
    \item \textbf{Blind Information Acquisition}: As demonstrated in recent studies \cite{shi2023large,asai2023self}, this compulsive retrieval mechanism without proper consideration of necessity or relevance can introduce irrelevant contextual information that undermines the MLLM's capability for accurate responses. Moreover, it neglects the inherent capabilities of MLLMs to reason and process multimodal data, rendering the retrieval step redundant or counterproductive.
    \item \textbf{Inadequate Query Formation}: The visual incompleteness, textual ambiguity, and conciseness create a fundamental impediment that fails to retrieve pertinent information. Furthermore, these existing single-step MRAG methodologies prove inadequate for addressing multi-hop queries that require a multi-step reasoning process \cite{kil2024ii}.
\end{itemize}

In response to these real-world limitations, we propose a new task, Multimodal Retrieval-Augmented Generation Planning (MRAG Planning). The objective is to establish optimized trajectories for information seeking and integration in multimodal queries. It guies the downstream MLLMs toward more accurate and comprehensive responses, while reducing the computational overhead. This task constitutes a central component in the development of effective agentic MRAG systems.
To this end, the MRAG Planning task is decomposed into two interrelated sub-tasks: 
(1) Information acquisition that discerns the truly necessary information required by the MLLMs and devises the appropriate retrieval strategies accordingly. 
(2) Query reformulation, which involves decomposing complex, multi-hop queries into manageable atomic sub-queries and refining them for clarity and informativeness. 
Figure \ref{fig:intro} illustrates a multi-hop reasoning example where the first step is to identify the game name AstroBot through an image search. The initial query is then decomposed into two sub-queries, each focusing on the sales data for the two respective games. Subsequent web searches are conducted to retrieve sales data and generate the final response. This example underscores the crucial need for dynamic planning procedures in MRAG systems where both order and selection of retrieval methods, and query reformulation are determined tailored to the specific characteristics of the multimodal query.

To address the challenges of MRAG Planning, we propose a novel framework, CogPlanner \footnote{We name our framework as CogPlanner because of the inherent cognitive process of humans.}, inspired by human cognitive processes. 
Just as humans synthesize and gather multimodal information to address complex queries, adapting their reasoning based on prior knowledge, CogPlanner emulates this behavior through a centralized planning expert that dynamically determines a planning procedure in coordination with downstream MLLMs. 
CogPlanner operates through two core operations: query reformulation and retrieval action selection. Query reformulation involves breaking down a complex query into related sub-queries or refining the queries. The retrieval strategy, in turn, encompasses image search, text search, or none. When sufficient information is gathered, the framework refrains from further retrieval, culminating in the generation of a final response. This iterative procedure reflects an adaptive chain of actions tailored to the specific needs of each multimodal query. For instance, the optimal planning strategy in Figure \ref{fig:intro} follows \textit{(Query Refinement, Image Search) $\rightarrow$ (Query Decomposition, Text Search)}.
CogPlanner supports two distinct modeling approaches: parallel and sequential modeling. Each differs in the order of the query reformulation and retrieval action selection.

In conjunction with the introduction of the CogPlanner framework, we present CogBench, a comprehensive dataset tailored to the MRAG Planning task. CogBench consists of over 5,000 data samples, with a high-quality test set of 401 samples. 
In addition to its essential role in evaluation, the development of CogBench facilitates the design of specialized fine-tuning strategies aimed at bolstering the decision-making capabilities of smaller, resource-efficient MLLMs. By utilizing the CogBench training set, we achieve lightweight integration of the Qwen2-VL-7B-Instruct \cite{Qwen2VL} model as the planning expert in CogPlanner. This integration, referred to as Qwen2-7B-VL-Cog, maintains its resource-efficient characteristics while enabling effective performance within the MRAG Planning context.
In summary, the contributions of this work are as follows:
\begin{itemize}[leftmargin=*]
    \item We thoroughly examine the limitations of current MRAG frameworks, specifically addressing the challenges of information acquisition and query reformulation. Building upon this, we formally define the task of Multimodal Retrieval Augmented Generation Planning (MRAG Planning), laying the groundwork for further research.
    \item We introduce CogPlanner, a flexible, plug-and-play framework that incorporates two distinct modeling approaches, parallel modeling and sequential modeling. 
    \item We develop the CogBench benchmark, tailored to the MRAG Planning task. It supports performance evaluation and facilitates fine-tuning strategies to enable the lightweight integration of resource-efficient MLLMs with CogPlanner. Experimental results demonstrate that CogPlanner achieves more than 15\% improvements over various MRAG approaches while incurring less than 10\% additional costs with Qwen2-VL-7B.
\end{itemize}

\section{Related Work}
\subsection{Query Processing in IR}
Query processing is a critical aspect of Information Retrieval (IR) systems, directly influencing the efficiency and effectiveness with which relevant information is retrieved in response to user queries. Early IR systems rely on complex, multi-stage query processing pipelines, which incorporate a range of techniques, including query rewriting \cite{query_rewrite}, intention detection \cite{arora2024intent}, sentiment analysis \cite{zhang2023sentiment}, and query expansion \cite{query2doc}, among others. These pipelines often utilize human-defined heuristics to refine the query, enabling more precise document retrieval \cite{li_ragsurvey, lee2020learning}. However, the advent of large language models (LLMs) has significantly transformed this approach. The exceptional expressive power and reasoning capabilities of LLMs allow them to effectively perform several traditional query processing tasks within a single, well-crafted prompt. This shift has been particularly notable in the context of RAG systems, where the primary challenge now lies in determining the most effective strategy for processing user queries.

\subsection{Multimodal Retrieval Augmented Generation}
RAG frameworks have demonstrated considerable success in various real-world applications \cite{li_ragsurvey}. However, their reliance on textual information presents a significant limitation, as it precludes the incorporation of crucial knowledge embedded within other modalities, such as images and videos. Multimodal Retrieval Augmented Generation (MRAG) seeks to address this limitation by equipping MLLMs with access to a broader spectrum of knowledge, encompassing up-to-date and domain-specific information \cite{zhao2023retrieving}. Empirical studies consistently demonstrate the effectiveness of MRAG systems \cite{mrag_survey} across various visual question answering (VQA) benchmarks \cite{vqa_survey}. 
Recent advancements in MRAG have demonstrated notable progress. For instance, MuRAG \cite{chen2022murag} highlights how incorporating visual information retrieved from external sources significantly improves the system performance. Other prominent approaches include Plug-and-Play \cite{plugandplay}, which transforms visual content into textual descriptions to facilitate integration with conventional text-based question-answering mechanisms. Additionally, RAMM \cite{yuan2023ramm} enhances the generation process by incorporating both text-to-image retrieval and subsequent fusion of the representations for more accurate answer generation. Further innovations include Wiki-llava \cite{wikillava} and mR2AG \cite{zhang2024mr2ag}, which enable retrieval from online knowledge bases, such as Wikipedia, via image-based queries, to provide more contextually informed responses to user queries. M2RAG \cite{m2rag} extends these efforts by enabling concurrent retrieval of both textual and visual elements in response to multimodal queries, allowing for more robust query understanding and generation capabilities. Additionally, benchmarks such as MRAG-Bench \cite{hu2024mrag} and MMSearch \cite{jiang2024mmsearch} have been introduced to evaluate MRAG performance, particularly in tasks with image-to-image retrieval, addressing challenges related to incomplete or insufficient image data.
The prevailing methodologies predominantly adhere to a rigid, single-modality search paradigm. However, in authentic user scenarios, knowledge acquisition can originate from diverse sources, contingent upon the specific query and the underlying domain. To this end, we introduce the novel task of MRAG Planning. The primary goal of MRAG Planning is to systematically determine the most effective query processing strategy tailored to each multimodal query and the underlying MLLMs.

\section{Task Formulation}
Consider a multimodal query $\mathcal{Q}_0 = (q, v)$ where $q$ represents the textual component and $v$ represents the visual component (e.g., an image). The objective of MRAG is to retrieve pertinent information from a document collection $\mathcal{D} = \{\mathcal{D}_1, \dots, \mathcal{D}_n\}$, and generate cogent responses. 
Drawing parallels with human information processing, we introduce the task of MRAG Planning that interfaces intimately with the retrieval tools and downstream MLLMs in the MRAG systems, restructuring their information gathering mechanism.
We formalize the MRAG system environment as a tuple $(\mathcal{G}, \mathcal{I})$. Here, $\mathcal{G}$ refers to the goal conditions of assembling sufficient information to generate comprehensive and accurate responses. $\mathcal{S}$ represents the state, capturing the current set of information available, which may encompass queries and any retrieved documents. The initial state corresponds to the input multimodal query, $\mathcal{I} = \mathcal{S}_0 = \mathcal{Q}_0$. The MRAG Planning task can thus be framed as a state transition function $\mathcal{F}$ that progresses from $\mathcal{I}$ toward the goal state through a chain of decisions. Formally, this transition process is defined as follows:
\begin{align}
    \mathcal{F}: \mathcal{S} \times \mathcal{P} \rightarrow \mathcal{S},
\end{align}
where $\mathcal{P}$ represents the available decision space.

\subsection{Multimodal Retrieval Augmented Generation Planning}
\label{formulation}
In line with the human cognitive architecture's capacity for knowledge integration, we conceptualize the MRAG Planning task as a dual optimization problem comprising two sub-tasks: information acquisition and query reformulation.

\subsubsection{Planning Procedure}
The decision space can thus be decomposed as $\mathcal{P} = (\mathcal{A}, \mathcal{Q})$, where $\mathcal{A}$ represents the information acquisition strategy and $\mathcal{Q}$ denotes the query reformulation result. To accommodate multi-hop reasoning queries, the planning process unfolds iteratively across $T$ rounds under the assumption of Markovian state transitions. Specifically, at each iteration $t$, the decision is determined by the current available information state:
\begin{align}
    \mathcal{S}_{t} \rightarrow (\mathcal{A}_{t}, \mathcal{Q}_{t}).
\label{decision}
\end{align}
The subsequent state is updated naturally as follows:
\begin{align}
    \{\mathcal{S}_t, (\mathcal{A}_{t}, \mathcal{Q}_{t})\} \rightarrow \mathcal{S}_{t+1},
\label{state_update}
\end{align}
where $\mathcal{S}_t$ may encompass the historical queries or the retrieved document elements, denoted as $\mathcal{D}^t$.

\subsubsection{Information Aquisition}
Recognizing the inherent limitations of MLLMs in terms of specific knowledge gaps, an information acquisition mechanism is imperative to supplement the MLLM's knowledge base. We define $\mathcal{A}$ as comprising three distinct retrieval operations: text search, image search, and non-search. The optimal retrieval action is determined based on the quality of the available multimodal information and the estimated utility of additional context from external knowledge sources:
\begin{align}
    \mathcal{A}_{t} = \argmax_{a \in \mathcal{A}} \mathcal{F}_{IA}(a \vert \mathcal{S}_t).
\end{align}
Post action selection, we proceed with in-document retrieval, which identifies and extracts pertinent elements under the selected retrieval strategy, yielding retrieved document elements $\mathcal{D}^t$.

\subsubsection{Query Reformalization}
To address the ambiguity and potential incompleteness in queries, we refine the queries by leveraging both textual and visual cues within the multimodal query. For complex queries necessitating multi-hop reasoning, we employ a decomposition strategy that preserves semantic relationships while breaking down the query into manageable sub-queries. This process can be expressed as:
\begin{align}
    \mathcal{Q}_{t} = \mathcal{F}_{QR}(\mathcal{Q}_{t-1}, \mathcal{S}_t) = \{\mathcal{Q}_{t,1}, \ldots, \mathcal{Q}_{t,N_t}\},
\end{align}
where $N_t$ represents the cardinality of the decomposed and refined query set at iteration $t$.

\subsubsection{Generation}
Once the final state $\mathcal{S}_T$ is deemed sufficiently informed, the planning process culminates in response generation. The response is synthesized by incorporating the initial query, the final refined query, and the relevant document elements:
\begin{equation}
\text{Response} = \mathcal{F}_{Gen}(\mathcal{Q}_0, \mathcal{Q}_{T}, \mathcal{D}^{T}),
\end{equation}
where $\mathcal{F}_{Gen}$ represents the MLLM generator.

\begin{figure*}
    \centering
    \includegraphics[width=1.0\linewidth]{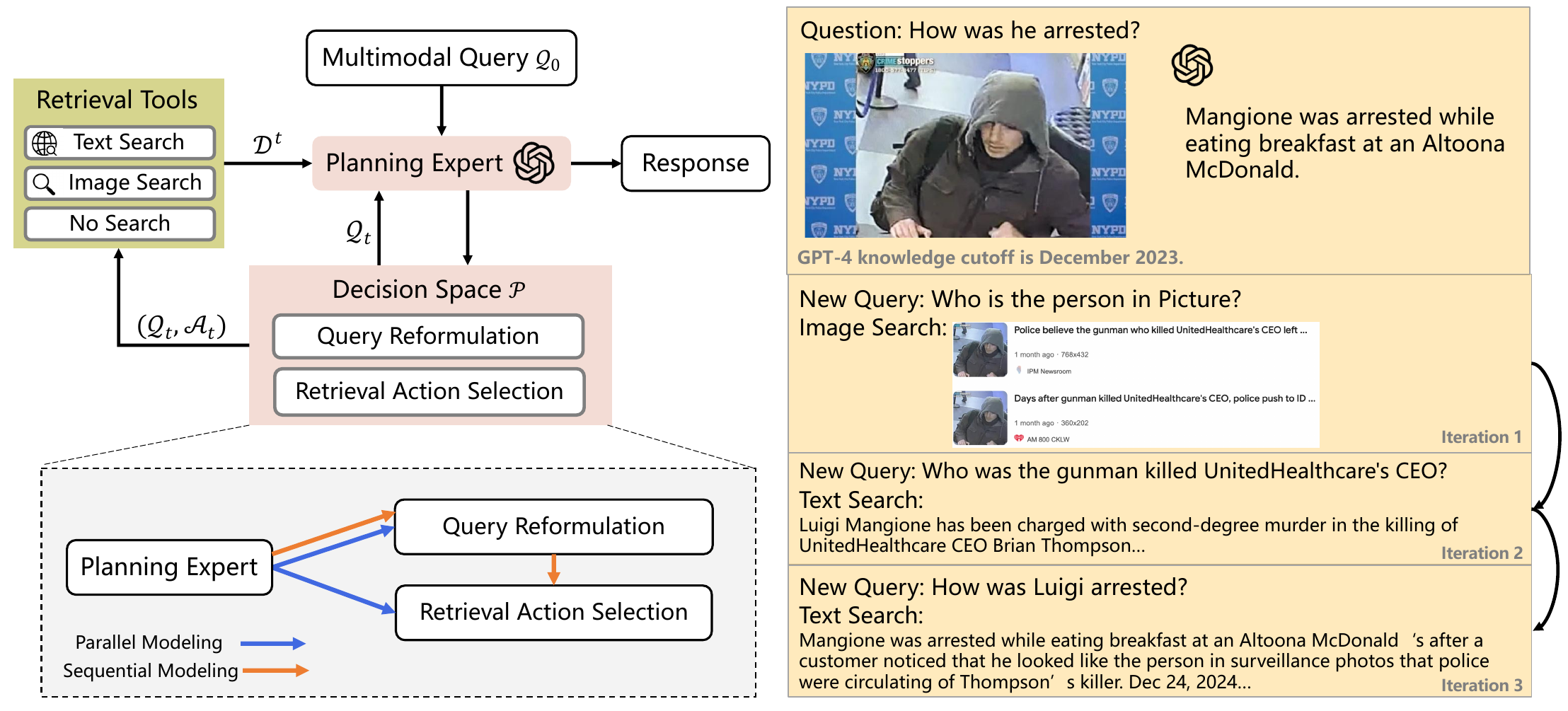}
    \caption{The overall framework of CogPlanner.}
    \label{framework}
\end{figure*}

\section{Methodology}
\subsection{Baselines}
Existing MRAG methodologies predominantly rely on a fixed information acquisition pipeline, characterized by a single-modality retrieval action performed in a single turn. These methods can be broadly classified into two distinct categories: 
\begin{itemize}[leftmargin=*]
    \item Fixed textual retrieval, which closely resembles traditional RAG frameworks. They employ textual queries to retrieve relevant documents to generate the final answer. 
    \item Fixed visual retrieval, which prioritizes visual information, leveraging visual queries to retrieve relevant images and their associated captions. The MLLM then integrates both the original user query and the augmented retrieved images to produce the final answer.
\end{itemize}
While these fixed workflows have been effective within certain domains, they exhibit limited flexibility. The rigid, modality-specific structure hinders their ability to adapt to other contexts, or the non-search scenario, where fixed search may bring extra noise. It undermines their potential for broader applicability in dynamic, real-world scenarios. 

\subsection{CogPlanner}
\label{mindplanner}
We propose CogPlanner, a flexible, plug-and-play framework that mirrors the human cognitive processes when handling complex multimodal queries. 
Our approach is inspired by the observation that humans demonstrate both adaptability and efficiency in acquiring and integrating multimodal information. When confronted with such queries, individuals instinctively engage in a structured, multi-step process that involves continuous assessment of information gaps, determination of appropriate retrieval strategies for the missing knowledge, and decomposition and refinement of complex queries into manageable sub-components. Crucially, this process is inherently adaptive - guided by the individual’s prior knowledge and cognitive capabilities, with the goal of converging toward a state of conceptual clarity and informational completeness.


CogPlanner operationalizes this cognitive architecture through an iterative decision-making framework. It dynamically orchestrates a chain of decisions for each query, optimizing both effectiveness and efficiency in conjunction with the downstream MLLMs. As shown in Figure \ref{framework}, the multimodal query — seeking up-to-date news about Luigi's arrest — requires three processing rounds to gather sufficient information for the current GPT knowledge base. CogPlanner centers on two critical decisions in $\mathcal{P}$: (1) query reformulation, and (2) action selection among text search, image search, and non-search, corresponding to the sub-tasks outlined in Section \ref{formulation}. We implement the core state transition function $\mathcal{F}$ through a planning expert who makes these decisions at each iteration, completing the roles of $\mathcal{F}_{IA}$ and $\mathcal{F}_{QR}$. Then, the retrieval is invoked to assess and process relevant multimodal elements.

\subsubsection{Planning Expert}
We employ an MLLM as the planning expert. As formalized in Equation \ref{decision},  at each iterative step, the expert analyzes the current multimodal content, evaluates the information gathered thus far, and subsequently determines the most appropriate follow-up retrieval action and reformulates the query. To accomplish this, we propose two distinct modeling paradigms - parallel modeling and sequential modeling - each of which differs in the order in which the decision-making occurs. \\

\noindent\emph{\textbf{Parallel Modeling}} 
In the parallel modeling paradigm, the planning expert concurrently adjusts the query and determines the appropriate retrieval action. Specifically, it takes the current query alongside the information retrieved in the preceding iteration as inputs. The implementation employs two parallel threads of MLLM inference: one is responsible for query reformulation, while the other determines the next retrieval action. The primary advantage of this paradigm lies in efficiency. By enabling simultaneous decision-making, we achieve a streamlined decision chain and faster processing. This is particularly beneficial in real-world applications where response latency is a critical consideration. The parallel process can be formally expressed as:
\begin{equation}
\begin{aligned}
    \mathcal{Q}_t =& \mathcal{F}_{QR}(\mathcal{Q}_{t-1}, \mathcal{D}^{t}), \\
    \mathcal{A}_t =& \mathcal{F}_{IA}(\mathcal{Q}_{t-1}, \mathcal{D}^t).
\end{aligned}
\end{equation} \\

\noindent\emph{\textbf{Sequential Modeling}} 
The sequential paradigm, in contrast, implements an ordered two-step decision process. The first step involves query reformulation, wherein multimodal inputs, along with previously retrieved information, are leveraged to refine and decompose the query. These restructured queries then serve as the inputs for the subsequent stage, which entails an evaluation of the necessity of further retrieval actions. This paradigm facilitates a more nuanced understanding of information retrieval requirements, as the planning expert is presented with both the original query and its reformulated meta-queries. It mimics a reflective cognitive process where the planning expert jointly assesses whether the queries align with the system’s knowledge bounds. Such an assessment enables the expert to more accurately determine whether additional retrieval actions are needed. This capability is especially beneficial for complex queries that demand deeper reasoning regarding information sufficiency. The sequential modeling process is formalized as follows:
\begin{equation}
\begin{aligned}
    \mathcal{Q}_t =& \mathcal{F}_{QR}(\mathcal{Q}_{t-1}, \mathcal{D}^{t}), \\
    \mathcal{A}_t =& \mathcal{F}_{IA}(\mathcal{Q}_{t}, \mathcal{D}^t).
\end{aligned}
\end{equation}

Both the retrieval decision-making function $\mathcal{F}_{IA}$ and query reformulation function $\mathcal{F}_{QR}$ are implemented through MLLM generation. To effectively harness the capabilities of MLLMs within these planning sub-tasks, we meticulously engineer tailored prompts.

\subsubsection{Retrieval and Generation}
At each iterative stage, the selection of retrieval actions determines whether or not and which retrieval API is invoked, either text retrieval or image retrieval. We leverage Google Web Search and Google Image Search as our primary retrieval API service. 
For each text retrieval request, we retrieve the top-k search results to ensure that only the most relevant information is retained. These results undergo preprocessing through the Jina API framework \footnote{https://github.com/jina-ai/reader}, which transforms the raw web content into structured representations better suited for MLLM consumption. 
The visual retrieval pipeline captures full-page screenshots of search results, employing a set of systematic human-crafted rules to eliminate extraneous elements such as white space and original query images. To balance computational cost with retrieval quality, we limit the image retrieval to between three and six high-confidence candidates, accompanied by their contextual captions to provide relevant semantic grounding.
Furthermore, to enhance system efficiency and responsiveness, we incorporate caching mechanisms for both text and image retrieval modules.

This iterative cycle culminates when the planning expert collectively assesses that the acquired information is sufficiently comprehensive and the formulated query exhibits adequate clarity. Upon reaching this convergence criterion, the planning procedure is finalized, and CogPlanner proceeds to generate the ultimate response.

\subsection{Compatibility}
The CogPlanner framework is inherently agnostic to the specific model employed, making it easy to be integrated into any MRAG system and immediately enhancing their performance, demonstrating stunning flexibility in real-world applications. The planning expert responsible for query reformulation and determining the appropriate retrieval action can be any MLLM, or even a traditional classification model.
For the planning expert, we exclusively employ a diverse set of advanced MLLMs as the foundation. Specifically, we leverage both closed-source APIs and open-source MLLMs. The closed-source models include GPT-4o \cite{hurst2024gpt}, while the open-source models consist of the Qwen-VL series \cite{Qwen-VL} and the Pixtral series \cite{dubey2024llama}. 

\section{CogBench Construction}
In this section, we present CogBench, a benchmark specifically developed for the MRAG planning task. CogBench comprises over 5,000 data samples, with a high-quality test set containing more than 400 samples. This benchmark is designed to facilitate the assessment of the effectiveness of our proposed CogPlanner framework, as well as other MRAG planning frameworks. Moreover, CogBench can be leveraged to enhance the decision-making capabilities of various MLLMs, particularly resource-efficient models, through fine-tuning.
In the following subsections, we detail the construction process of CogBench and demonstrate how the benchmark enables lightweight integration of CogPlanner with the Qwen2-VL-7B-Instruct.

\subsection{Query Collection}
The rapid evolution of MLLMs has underscored the need for evaluation on increasingly complex user queries that mirror real-world application scenarios. While existing benchmarks \cite{jiang2024mmsearch,hu2024mrag} provide valuable groundwork, we recognized the necessity to extend beyond their scope. To this end, we deliberately incorporate more complex queries that demand image-based knowledge augmentation. We acquire authentic user intent through web-sourced screenshots. We curate a diverse array of topics and structure search queries around these topic words, such as \textit{"Astro Bot screenshot"}, and use Google Image Search to collect an image corpus. The resultant image corpus is subject to a manual filtration process. To generate realistic queries, we leveraged the Claude-3.5-sonnet API \footnote{https://www.anthropic.com/news/claude-3-5-sonnet} to simulate real users, producing five distinct queries per image that span both factual and open-ended inquiries requiring visual context interpretation. Each query-image pair undergoes manual review and modification by two senior AI research engineers, each bringing at least three years of domain expertise. The modification process follows several key principles: (1) each query must be distinct, with no repetition—even across different images; (2) queries must be unambiguous; (3) queries should be meaningful and formulated naturally, resembling how real humans would ask them; and (4) each query should target specific information, asking about concrete aspects of the image. For each image, the 1–5 most compelling queries, which highlight the potential of multimodal retrieval, are retained.

\subsection{MRAG Planning and Generation}
The MRAG planning and generation process is central to the CogPlanner framework, as detailed in Section \ref{mindplanner}. We employ the GPT-4o API for executing the planning process. This implementation records each iteration of the planning process, encompassing the series of actions taken, the multimodal document sets retrieved, and the responses generated.
Following response generation, the expert annotators conduct a thorough examination to review and regularize the entire chain of actions, and manually annotate the golden answer. Each data sample in the CogBench, therefore, contains a multimodal query, the retrieval actions, reformulated queries at each iterative step, the documents retrieved, and the final golden answer.
To be noticed, we do not define a fixed gold standard for the multimodal query processing, as manual annotation of information collection paths does not yield a unique or definitive reference. Instead, we focus on the correctness and completeness of the final answer.
Finally, the CogBench dataset is divided into training and test sets, comprising 5307 and 401 samples, respectively. 

\subsection{CogBench Analysis}
\begin{figure}[t]
    \centering
    \includegraphics[width=0.5\linewidth]{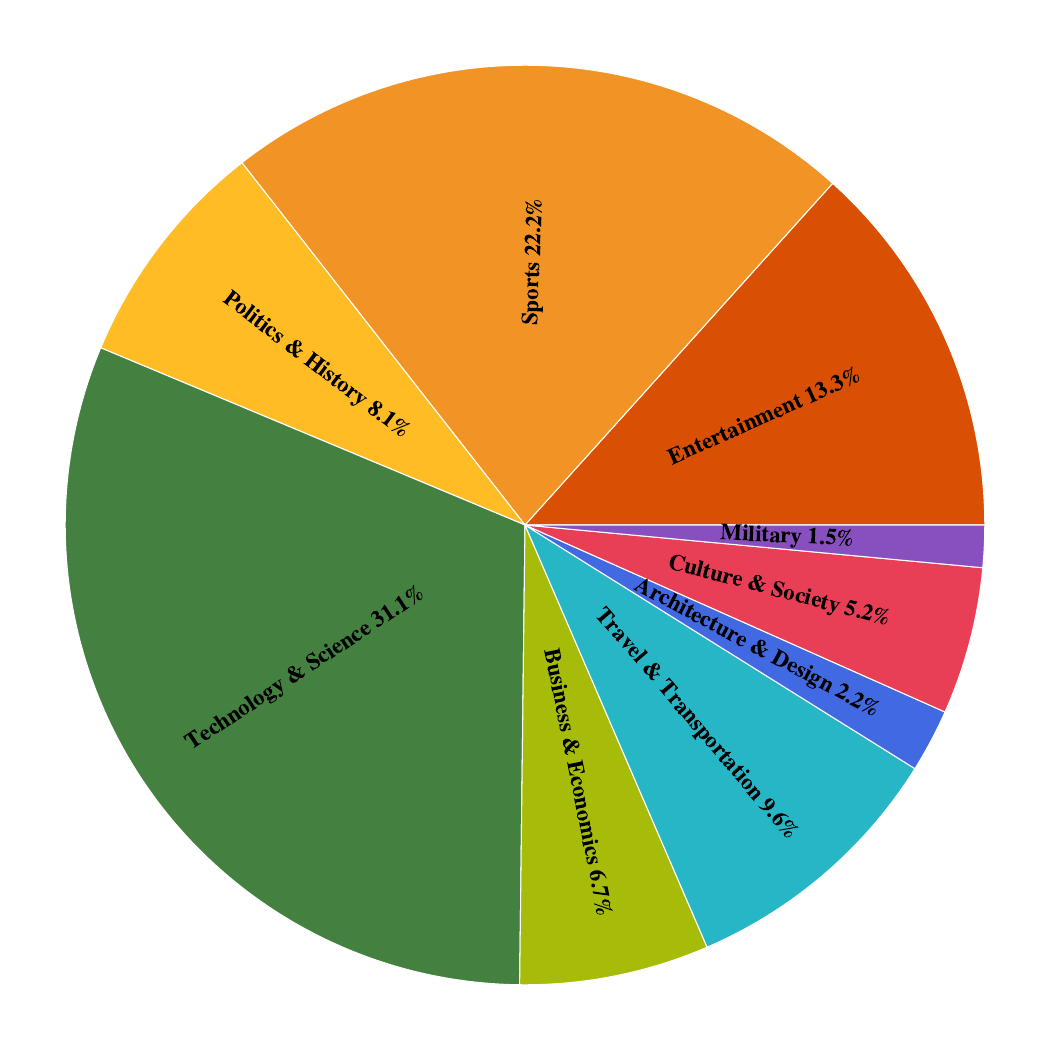}
    \caption{Domain distribution of CogBench.}
    \label{domain}
\end{figure}

As shown in Table \ref{stati} and Figure \ref{domain}, CogBench contains 5718 user queries spanning 9 distinct cognitive domains. We identify several fundamental characteristics that distinguish CogBench from existing benchmarks. 
(1) Unlike previous benchmarks, which focus primarily on query-response pairs, CogBench offers a comprehensive record of the entire planning procedure involved in MRAG tasks, thereby facilitating the training of MLLMs. 
(2) A critical limitation of current MRAG benchmarks lies in their reliance on artificial query construction, primarily through simple entity substitution techniques \cite{rqvqa,hu2024mrag}. Such methodologies typically yield responses confined to single entities or numerical values, severely understating the complexity inherent in real-world multimodal interaction scenarios. In contrast, CogBench introduces 24.19\% open-ended queries that demand sophisticated, multi-faceted responses encompassing multiple interconnected claims with much longer answer length - 40.13 tokens on average.
(3) CogBench incorporates diverse planning procedures that necessitate distinct search strategies at different stages, resulting in varied decision chains across different MLLMs. To quantify this complexity, we also ask the annotators to assess the number of reasoning steps required for resolution (e.g. a single round of question answering is considered a 1-hop query). Our findings reveal that 79.55\% of cases explicitly require MRAG Planning, highlighting the sophisticated nature of the reasoning tasks presented in our benchmark.

\begin{table}[t]
    \centering
    \caption{Key statistics of CogBench.}
    \setlength{\tabcolsep}{4pt} 
    \begin{tabular}{ccccc}  
    \toprule[1pt]  
     \# Query & \# Domians & \# Query Len. & \# Answer Len. & \# Images \\
    \midrule
    5718 & 9 & 8.95 & 40.13 & 1381 \\
    \midrule
    \multicolumn{3}{c}{Reasoning Steps} & \multicolumn{2}{c}{Answer Type} \\
    \cmidrule(lr){1-3} \cmidrule(lr){4-5} 
    1-hop & 2-hop & \multicolumn{1}{c}{> 2-hop} & open-ended & close-ended \\
    \cmidrule(lr){1-3} \cmidrule(lr){4-5} 
     1166 & 1882 & \multicolumn{1}{c}{2666} & 1383  & 4334 \\
     20.39\% & 32.91\% & \multicolumn{1}{c}{46.62\%} & 24.19\% & 75.80\% \\
    \bottomrule[1pt]
    \end{tabular} 
\label{stati}
\end{table}

\vspace{-2mm}

\subsection{Lightweight Integration of CogPlanner}
To improve the efficiency and reduce the resource requirements of the CogPlanner framework, we aim to achieve a more lightweight integration within MRAG systems. Building upon the CogBench, we introduce a specialized fine-tuning strategy tailored for smaller, resource-efficient MLLMs to broaden the applicability of the CogPlanner framework and mitigate resource constraints. We utilize the Qwen2-VL-7B-Instruct as the backbone and employ the CogBench training set as specialized training data. To maintain a balance between the fine-tuning process and the retention of the Qwen2-VL-7B-Instruct model's general capabilities, we augment the training dataset with general instruction data at a 1:1 ratio. This ensures that the model benefits from the specialized training required for CogPlanner integration while preserving its broad functionality. The fine-tuned model is referred to as Qwen2-7B-VL-Cog. 
This fine-tuning methodology is highly adaptable and capable of being applied to any existing MLLM. In our practice, we find that the CogBench fine-tuning process significantly enhances the MRAG planning capabilities of MLLMs, making them qualified for effective planning experts. Ultimately, this approach facilitates the development of a lightweight integration of CogPlanner, enhancing MRAG performance while requiring minimal additional computational resources.

\begin{table*}
\setlength{\tabcolsep}{4pt} 
\renewcommand{\arraystretch}{1} 
\caption{Performance comparison between CogPlanner and baseline MRAG methodologies on CogBench. Precision and recall are evaluated at the claim level, while the F1 score is assessed at the token level. The diverse planning procedures required by CogBench lead to performance degradation across all fixed pipeline baselines, whereas CogPlanner demonstrates substantial improvements.}
\centering
\begin{tabular}{l|ccc|ccc|ccc|ccc}
\toprule[1pt]
\multirow{3}{*}{\textbf{Model}} & \multicolumn{9}{c|}{\textbf{Reasoning-Steps}} & \multicolumn{3}{c}{\textbf{Overall Performance}} \\ \cmidrule{2-13}
& \multicolumn{3}{c|}{\textbf{1-hop}} & \multicolumn{3}{c|}{\textbf{2-hop}} & \multicolumn{3}{c|}{\textbf{> 2-hop}}   \\ \cmidrule{2-13}
& Precision & Recall & F1 & Precision & Recall & F1 & Precision & Recall & F1 & Precision & Recall & F1   \\
\midrule
\multicolumn{13}{c}{\textbf{Origin MLLMs}} \\
\midrule
GPT-4o & 33.49 & 37.44 & 10.01 & 44.38 & 59.46 & 39.84 & 16.17 & 24.74 & 12.83 & 29.07 & 38.85 & 21.21 \\
Pixtral-Large-Instruct & 24.30 & 41.87 & 5.19 & 37.65 & 54.64 & 34.51 & 10.77 & 24.50 & 7.13 & 22.45 & 38.05 & 15.81 \\
QVQ-72B-Preview & 34.15 & 22.10 & 10.56 & 40.51 & 31.29 & 21.48 & 19.42 & 13.99 & 6.37 & 29.43 & 21.39 & 12.24 \\
Qwen2-VL-72B-Instruct & 34.45 & 48.85 & 7.95 & 35.46 & 44.94 & 37.00 & 12.55 & 17.00 & 9.27 & 24.63 & 32.78 & 18.19 \\
\midrule
\multicolumn{13}{c}{\textbf{With Fixed Image Retrieval}} \\
\midrule
GPT-4o & 31.73 & 23.13 & 10.74 & 36.75 & 45.14 & 36.07 & 17.88 & 30.10 & 11.13 & 26.97 & 33.67 & 19.32 \\
Pixtral-Large-Instruct & 23.10 & 39.27 & 6.13 & 28.05 & 44.51 & 29.28 & 12.58 & 24.91 & 6.88 & 19.86 & 34.35 & 14.16 \\
QVQ-72B-Preview & 11.35 & 23.13 & 2.72 & 17.10 & 19.77 & 20.42 & 8.17 & 10.24 & 2.70 & 11.78 & 16.03 & 8.58 \\
Qwen2-VL-72B-Instruct  & 26.30 & 35.65 & 9.02 & 26.92 & 32.44 & 29.20 & 16.45 & 17.45 & 7.89 & 21.93 & 26.14 & 15.19 \\
\midrule
\multicolumn{13}{c}{\textbf{With Fixed Text Retrieval}} \\
\midrule
GPT-4o & 22.68 & 15.63 & 5.72 & 38.72 & 36.79 & 37.08 & 15.72 & 17.11 & 10.49 & 24.77 & 23.33 & 18.33 \\
Pixtral-Large-Instruct & 8.35 & 14.22 & 1.57 & 16.91 & 21.73 & 27.27 & 8.58 & 13.01 & 7.18 & 11.30 & 16.15 & 12.70 \\
QVQ-72B-Preview & 29.22 & 18.54 & 8.04 & 24.90 & 16.62 & 18.75 & 9.95 & 8.08 & 6.02 & 18.85 & 13.05 & 10.66 \\
Qwen2-VL-72B-Instruct & 20.56 & 24.47 & 7.12 & 26.66 & 27.07 & 30.64 & 11.63 & 13.33 & 9.98 & 18.44 & 20.17 & 16.25 \\
\midrule
\multicolumn{13}{c}{\textbf{Self-Reflective RAG}} \\
\midrule
GPT-4o & 43.83 & 11.72 & 15.56 & 28.79 & 22.21 & 30.88 & 22.79 & 20.68 & 13.03 & 29.59 & 19.13 & 19.47 \\
Qwen2-VL-72B-Instruct & 43.29 & 10.97 & 10.49 & 27.55 & 20.72 & 28.58 & 19.01 & 18.96  & 13.85 & 26.81 & 17.91 & 18.05 \\
\midrule
\multicolumn{13}{c}{\textbf{CogPlanner with Parallel Modeling}} \\
\midrule
GPT-4o & 35.74 & 37.67 & 10.46 & 47.46 & 57.37 & 42.83 & 24.09 & 27.73 & 13.65 & 34.22 & 39.59 & 22.67  \\
Pixtral-Large-Instruct & 21.46 & 37.08 & 5.16 & 45.35 & 52.29 & 38.96 & 20.95 & 30.34 & 11.12 & 29.15 & 39.00 & 19.13 \\
QVQ-72B-Preview & 36.38 & 36.44 & 48.16 & 43.90 & 29.60 & 24.43 & 22.19 & 25.22 & 26.69 & 32.29 & 28.97 & 30.33 \\
Qwen2-VL-72B-Instruct & 35.50 & 48.77 & 8.23 & 45.88 & 47.08 & 38.51 & 21.64 & 24.00 & 15.71 & 32.45 & 36.90 & 21.74 \\
\midrule
\multicolumn{13}{c}{\textbf{CogPlanner with Sequential Modeling}} \\
\midrule
GPT-4o & 32.92 & 33.46 & 10.14 & 49.67 & 54.68 & 43.26 & 28.03 & 33.38 & 15.26 & \textbf{36.21} & \textbf{40.46} & 23.49 \\
Pixtral-Large-Instruct & 22.60 & 36.39 & 5.57 & 37.18 & 54.32 & 39.60 & 9.96 & 29.96 & 11.28 & 21.57 & 39.36 & 19.51 \\
QVQ-72B-Preview & 35.62 & 45.95 & 51.69 & 43.36 & 44.07 & 25.37 & 19.54 & 21.18 & 27.27 & 30.73 & 33.84 & \textbf{31.63} \\
Qwen2-VL-72B-Instruct & 36.88 & 48.86 & 7.84 & 42.94 & 44.45 & 37.48 & 21.57 & 25.01 & 14.27 & 31.79 & 36.33 & 20.65 \\
\bottomrule[1pt]
\end{tabular}
\label{main}
\end{table*}

\section{Experiments}
\subsection{Experimental Settings}
Our experimental evaluation of CogPlanner, conducted on the CogBench test set, encompasses two primary dimensions: the overall performance of the MARG system and the analysis of the planning procedure within CogPlanner. We utilize the following backbone MLLMs, GPT-4o \cite{hurst2024gpt}, Qwen2-VL-72B-Instruct \cite{Qwen2VL}, Pixtral-Large-Instruct \cite{jiang2024mixtral}, and our fine-tuned Qwen2-7B-VL-Cog. Notably, QVQ-72B-Preview serves as a representative MLLM for advanced multimodal reasoning capabilities.

\subsubsection{End-to-End MRAG Performance}
We conduct six distinct experimental configurations across all selected MLLMs. The baseline configuration employs the original MLLMs, where multimodal queries are processed directly by the MLLM. We then examine two intermediate configurations: one incorporating fixed image retrieval based on visual query components, and another utilizing fixed text retrieval driven by textual query components. Besides, we employ the self-reflective RAG framework\footnote{https://github.com/langchain-ai/langgraph/tree/main/examples/rag} \cite{asai2023self} as a reflective and iterative competitive framework. The core evaluation focuses on both parallel and sequential modeling implementations of CogPlanner. 

For performance metrics, we adopt both token-level and claim-level evaluations, inspired by \cite{ru2024ragchecker}. Token-level evaluation is performed using the F1 score, measuring the overlap of common tokens between the generated response and the ground truth. Specifically, we use the NLTK tokenizer to segment the generated answers and ground truth. For claim-level evaluation, we utilize both precision and recall, calculated by first extracting claims from both the golden and generated answers using GPT-4o. Precision measures the proportion of correct claims within the generated responses, while recall evaluates the proportion of correct claims relative to the ground-truth answer claims.

\begin{table}
\centering
\caption{Performance of query reformulation across different MLLMs.}
\begin{tabular}{l|l|ccc}
\toprule
Category & Model & BLEU & Rouge & F1 \\
\midrule
Prompting & GPT-4o & 0.1629 & 0.4951 & 0.5375\\
\midrule
\multirow{3}{*}{Parallel} & GPT-4o & 0.1922 & 0.5266 & 0.5620 \\
& Pixtral-Large-Instruct & 0.1678 & 0.4614 & 0.5089 \\
& Qwen2-VL-72B-Instruct & 0.0907 & 0.4140 & 0.4472 \\
 \midrule
\multirow{3}{*}{Sequential} & GPT-4o & 0.1739 & 0.5050 & 0.5460 \\
& Pixtral-Large-Instruct & 0.1773 & 0.4707 & 0.5221 \\
& Qwen2-VL-72B-Instruct & 0.0918 & 0.4266 & 0.4643 \\
\bottomrule
\end{tabular}
\label{requery}
\end{table}

\subsubsection{Planning Procedure Performance}
In addition to the overall MRAG performance, we examine the efficiency of CogPlanner's planning procedure, with emphasis on query reformulation, by comparing its reformulated queries with those annotated by human experts. We evaluate three distinct approaches: parallel modeling, sequential modeling, and direct reformulating query through GPT-4o with optimized prompt engineering. The comparative analysis employs standard metrics, including BLEU \cite{papineni2002bleu}, ROUGE \cite{lin2004rouge}, and F1 scores to assess the quality and relevance of the reformulated query outputs against established ground truth.

\begin{table*}
\setlength{\tabcolsep}{8pt} 
\renewcommand{\arraystretch}{1} 
\caption{End-to-end performance and efficiency evaluation of Qwen2-VL-72B-Instruct and Qwen2-7B-VL-Cog as planning experts, with response generation models consistently using Qwen2-VL-72B-Instruct.}
\centering
\begin{tabular}{l|cccc|cc}
\toprule[1pt]
\textbf{Planning Expert} & \textbf{Precision} & \textbf{Recall} & \textbf{F1} & \textbf{Avg} & \textbf{\# Total Tokens} & \textbf{Latency(s)} \\
\midrule
\multicolumn{7}{c}{\textbf{CogPlanner With Parallel Modeling}} \\
\midrule
Qwen2-VL-72B-Instruct & 32.45 & 36.90 & 21.74 & 30.36 & 9.76 (14.9\%) & 1.209 \\
Qwen2-7B-VL-Cog & 31.97 & 32.65 & 21.46 & 28.69 & 7.58 (9.8\%) & 0.484 \\
\midrule
\multicolumn{7}{c}{\textbf{CogPlanner With Sequential Modeling}} \\
\midrule
Qwen2-VL-72B-Instruct & 31.79 & 36.33 & 20.65 & 29.59 & 13.56 (21.6\%) & 1.842  \\
Qwen2-7B-VL-Cog & 32.50 & 33.10 & 21.38 & 29.00 & 7.59 (11.9\%) & 0.545 \\
\bottomrule[1pt]
\end{tabular}
\label{light}
\end{table*}

\subsubsection{Implementation Details}
In our retrieval process, we retain the top five results from a web search, each result containing a maximum of 800 tokens. For fine-tuning the Qwen2-7B-VL-Cog model, we leverage the Llama-Factory framework \cite{zheng2024llamafactory} with a learning rate of 2e-6. 
A cosine learning rate scheduler is employed, and training proceeds for 2 epochs with a batch size of 32 and a warm-up ratio of 0.1. 
To ensure computational efficiency and prevent indefinite reasoning loops, we impose an upper bound of three iterations on the CogPlanner planning process.
Our experiments are conducted on 8 NVIDIA A800 GPUs.

\subsection{CogPlanner Performance}
\subsubsection{End-to-end Performance}
Table \ref{main} presents a comprehensive comparison of the end-to-end performance of various MLLMs integrated with current MRAG methodologies and our proposed CogPlanner. The following observations can be drawn based on these results:
(1) \textbf{Enhanced Performance with CogPlanner}: Notably, CogPlanner with GPT-4o consistently yields best performance compared to all other configurations. Specifically, it delivers substantial improvements over baseline MRAG systems, with end-to-end performance gains ranging from 12.4\% to 52.5\%, and at least a 41.45\% improvement over self-reflective MRAG variants. This enhancement is attributed to its ability to decompose and refine complex\ queries. By simplifying these queries, CogPlanner facilitates the dynamic determination of necessary retrieval actions, thereby ensuring the acquisition of accurate, complementary information. These results highlight the critical role of MRAG Planning in optimizing the performance of MRAG systems.
(2) \textbf{Weakness of Fixed Search Strategies}: The two fixed search strategies, while offering some improvements in specific metrics, generally exhibit a negative impact when compared to direct responses generated by MLLMs. As anticipated, these rigid search actions, particularly when applied to concise queries and screenshot images within multimodal queries, introduce noise that misleads downstream MLLMs rather than providing useful information. The noisy retrieval results tend to obscure the relevant information, thereby diminishing the effectiveness of the system. These findings further validate the motivation behind the development of the CogPlanner framework.
(3) \textbf{Comparison of Parallel and Sequential Modeling Approaches}: Comparing our parallel and sequential modeling methodologies, they show comparable performance, each demonstrating significant improvements over traditional MRAG systems. However, the sequential modeling approach does not exhibit a substantial advantage over the parallel approach. This suggests that the current capabilities of MLLMs are insufficient for accurately evaluating complementary information in a sequential manner
(4) \textbf{Benefits on Multihop Query}: The performance of CogPlanner shows variability across 1-3 hop query categories, with more pronounced improvements observed in multi-hop queries. This pattern suggests that the iterative and adaptive planning process of CogPlanner is particularly beneficial in scenarios that require the retrieval of information from multiple sources and involve multi-step reasoning to formulate a complete and accurate response.


\subsubsection{Query Reformulation Performance}
As illustrated in Table \ref{requery}, we evaluate the query reformulation performance of CogPlanner in comparison with various backbones, including direct prompt engineering approaches with these MLLMs. Among the configurations tested, GPT-4o with parallel modeling emerges as the highest-performing setting. Notably, CogPlanner demonstrates its ability to significantly enhance the accuracy and informativeness of the original multimodal queries, thereby improving the overall query formulation process. This improvement underscores the core objective of query reformulation within the context of MRAG Planning, highlighting the robustness and effectiveness of the CogPlanner framework. Moreover, the results reveal that the query reformulation performance is highly contingent upon the selection of the planning expert MLLMs.


\subsection{Lightweight Model Performance}
In this section, we assess the performance of our fine-tuned Qwen2-7B-VL-Cog from both effectiveness and efficiency perspectives. The results are presented in Table \ref{light}.

\subsubsection{Model Evaluation}
To ensure a fair comparison, we use the Qwen2-VL-72B-Instruct as the reference model for final answer generation. Both Qwen2-VL-72B-Instruct and our fine-tuned Qwen2-7B-VL-Cog are utilized as planning experts within the CogPlanner framework, facilitating an assessment of whether smaller MLLMs, specifically through our tailored supervised fine-tuning procedure, can effectively manage the complexities inherent in MRAG planning tasks.  The results underscore the critical role of model selection in determining the overall performance. While larger MLLMs typically demonstrate superior results, our Qwen2-7B-VL-Cog closely approximates the performance of Qwen2-VL-72B-Instruct across most evaluation metrics. This finding serves to validate the efficacy of our fine-tuning strategy. Under the CogPlanner framework, our results demonstrate that it is indeed possible to deploy a resource-efficient planning expert, leading to enhanced performance of the MRAG system. Specifically, the Qwen2-7B-VL-Cog emerges as a compelling alternative.

\subsubsection{Efficiency Evaluation}
We compare the efficiency of CogPlanner with Qwen2-VL-72B-Instruct and Qwen2-7B-VL-Cog as the planning expert. Specifically, we use total token generation and latency as the primary evaluation metrics, excluding the cost associated with final answer generation to isolate planning efficiency. As shown in Table \ref{light}, our findings indicate that Qwen2-7B-VL-Cog constitutes a significantly more lightweight module for MRAG systems, incurring only a 10\% increase in token consumption, but reducing latency to just 30\% compared to the Qwen2-VL-72B-Instruct model. It emerges as a practical compromise between performance and computational efficiency, particularly well-suited for deployment in real-world industrial settings. Furthermore, the parallel execution model exhibits superior performance relative to the sequential modeling approach across both efficiency metrics, aligning with our design expectations.

\subsection{Analysis}


\begin{table}[t]
\centering
\caption{The proportion of retrieval actions of different methodologies, \# No, \# Text, \# Image represents no search, text search and image search, respectively.}
\begin{tabular}{l|l|ccc}
\toprule
Model & Category & \# No & \# Text & \# Image \\
\midrule
\multirow{2}{*}{Pixtral-Large-Instruct} & Parallel & 11.51\% & 65.24\% & 23.25\%  \\
& Sequential & 8.43\% & 84.28\% & 18.00\% \\
\midrule
\multirow{2}{*}{Qwen2-VL-72B-Instruct} & Parallel & 13.38\% & 59.87\% & 26.75\%  \\
& Sequential & 11.46\% & 66.07\% & 22.47\% \\
\midrule
\multirow{2}{*}{Qwen2-7B-VL-Cog} & Parallel & 5.25\% & 80.43\% & 14.32\% \\
& Sequential & 5.25\% & 80.42\% & 14.43\% \\
\bottomrule
\end{tabular}
\label{planning}
\end{table}

\subsubsection{Analysis on planning procedure of Cogplanner}

\begin{figure}[!t]
  \centering
    \includegraphics[width=0.23\textwidth]{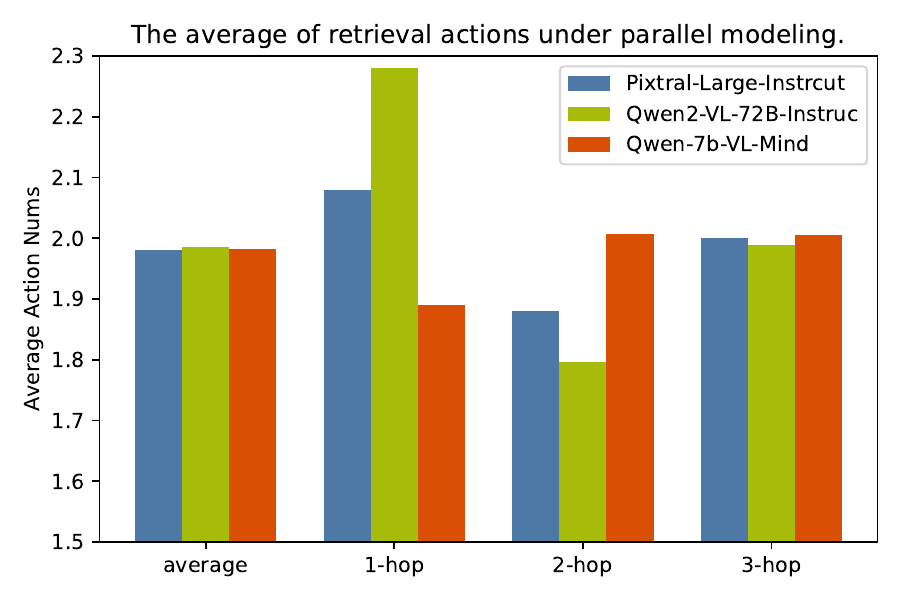}
    \includegraphics[width=0.23\textwidth]{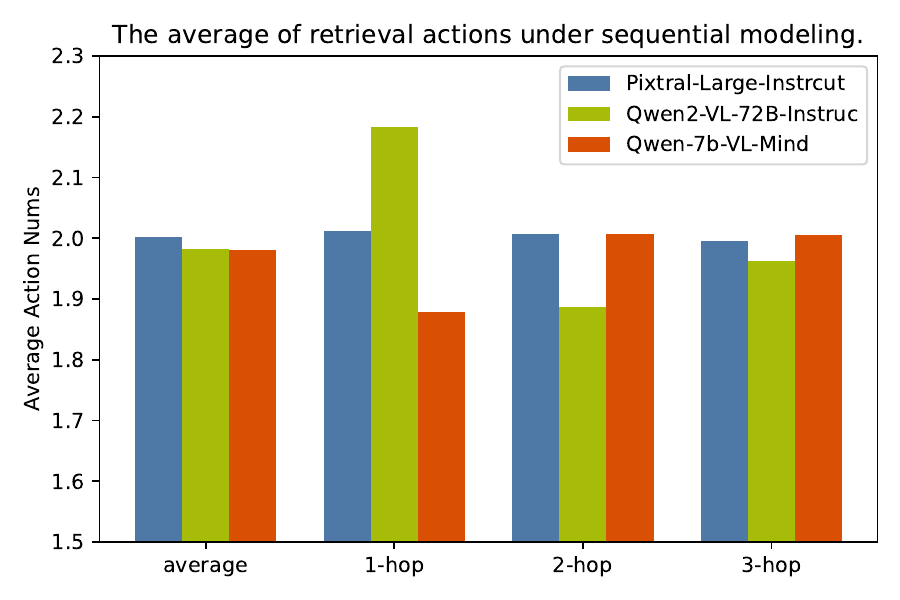}
\caption{The average number of retrieval actions of different MLLMs across queries with different reasoning steps.}
\label{planning_fig}
\end{figure}

We conduct an analysis to explore the adaptive decision-making capabilities of CogPlanner, which emulates human cognition by tailoring its planning processes to the specific knowledge bases of different MLLMs. Specifically, we examine the length of decision chains and the distribution of retrieval actions across the Pixtral-Large-Instruct, Qwen2-VL-72B-Instruct, and Qwen2-7B-VL-Cog. The results are summarized in Table \ref{planning} and Figure \ref{planning_fig}. The following observations can be drawn from these analyses:
(1) As shown in Figure \ref{planning_fig}, the distribution of retrieval actions indicates that all the MLLMs tend to perform more actions than are typically expected by human annotators. The expected behavior would be a gradual progression of actions. However, MLLMs generally opt to acquire more information and make conservative decisions in an attempt to ensure accuracy \cite{chen2024not,zhang2024path}. Among the models examined, the Qwen2-VL-72B-Instruct exhibits the most pronounced mismatch, performing over two rounds of processing even for 1-hop queries, which is counterintuitive when compared to its behavior on more complex, multi-hop queries. In contrast, the Qwen2-7B-VL-Cog model produces the most reasonable number of retrieval actions. Additionally, comparing sequential and parallel modeling paradigms reveals a substantial reduction in redundant retrieval actions. This is attributed to the query reformulation step, which allows the CogPlanner to better evaluate whether further search is genuinely necessary, resembling a reflective thought process.
(2) From the data presented in Table \ref{planning}, it is evident that the Qwen2-7B-VL-Cog model predominantly relies on textual search. This trend is also observable when comparing the two modeling paradigms. These results suggest that text-based retrieval remains the most preferred and fundamental method for acquiring information in MLLMs. Furthermore, the prevalence of redundant retrieval actions in this domain could be attributed to the tendency of models to perform additional web searches for reassurance, which does not necessarily harm overall performance.
(3) Notably, most tasks appear to be resolved around 2-hop retrieval steps, even for more complex queries requiring greater than 2 hops. Determining an optimal retrieval strategy that aligns with the model's knowledge base and reasoning capabilities remains a challenging task for current MLLMs.

\section{Conclusion}
This work introduces Multimodal Retrieval Augmented Generation Planning (MRAG Planning) to address the limitations of current MRAG frameworks. Our research underscores the importance of dynamically optimizing the information acquisition and query reformulation processes. The CogPlanner framework leverages decision-making to refine queries and select appropriate retrieval strategies, minimizing redundant retrieval and enhancing response quality. CogPlanner offers flexible modeling approaches and integrates seamlessly with existing MRAG systems. Additionally, we propose the CogBench benchmark to assess MRAG Planning’s decision-making capabilities, filling a gap in current evaluation methods. The experimental results validate the efficacy of MRAG Planning and CogPlanner, showcasing substantial improvements in performance with minimal additional computational costs. This work paves the way for more adaptive and effective MRAG frameworks.

\appendix

\bibliographystyle{ACM-Reference-Format}
\balance
\bibliography{sample-base}


\begin{thebibliography}{38}


\ifx \showCODEN    \undefined \def \showCODEN     #1{\unskip}     \fi
\ifx \showISBNx    \undefined \def \showISBNx     #1{\unskip}     \fi
\ifx \showISBNxiii \undefined \def \showISBNxiii  #1{\unskip}     \fi
\ifx \showISSN     \undefined \def \showISSN      #1{\unskip}     \fi
\ifx \showLCCN     \undefined \def \showLCCN      #1{\unskip}     \fi
\ifx \shownote     \undefined \def \shownote      #1{#1}          \fi
\ifx \showarticletitle \undefined \def \showarticletitle #1{#1}   \fi
\ifx \showURL      \undefined \def \showURL       {\relax}        \fi
\providecommand\bibfield[2]{#2}
\providecommand\bibinfo[2]{#2}
\providecommand\natexlab[1]{#1}
\providecommand\showeprint[2][]{arXiv:#2}

\bibitem[Arora et~al\mbox{.}(2024)]%
        {arora2024intent}
\bibfield{author}{\bibinfo{person}{Gaurav Arora}, \bibinfo{person}{Shreya Jain}, {and} \bibinfo{person}{Srujana Merugu}.} \bibinfo{year}{2024}\natexlab{}.
\newblock \showarticletitle{Intent Detection in the Age of LLMs}.
\newblock \bibinfo{journal}{\emph{arXiv preprint arXiv:2410.01627}} (\bibinfo{year}{2024}).
\newblock


\bibitem[Asai et~al\mbox{.}(2023)]%
        {asai2023self}
\bibfield{author}{\bibinfo{person}{Akari Asai}, \bibinfo{person}{Zeqiu Wu}, \bibinfo{person}{Yizhong Wang}, \bibinfo{person}{Avirup Sil}, {and} \bibinfo{person}{Hannaneh Hajishirzi}.} \bibinfo{year}{2023}\natexlab{}.
\newblock \showarticletitle{Self-rag: Learning to retrieve, generate, and critique through self-reflection}.
\newblock \bibinfo{journal}{\emph{arXiv preprint arXiv:2310.11511}} (\bibinfo{year}{2023}).
\newblock


\bibitem[Bai et~al\mbox{.}(2023)]%
        {Qwen-VL}
\bibfield{author}{\bibinfo{person}{Jinze Bai}, \bibinfo{person}{Shuai Bai}, \bibinfo{person}{Shusheng Yang}, \bibinfo{person}{Shijie Wang}, \bibinfo{person}{Sinan Tan}, \bibinfo{person}{Peng Wang}, \bibinfo{person}{Junyang Lin}, \bibinfo{person}{Chang Zhou}, {and} \bibinfo{person}{Jingren Zhou}.} \bibinfo{year}{2023}\natexlab{}.
\newblock \showarticletitle{Qwen-VL: A Versatile Vision-Language Model for Understanding, Localization, Text Reading, and Beyond}.
\newblock \bibinfo{journal}{\emph{arXiv preprint arXiv:2308.12966}} (\bibinfo{year}{2023}).
\newblock


\bibitem[Caffagni et~al\mbox{.}(2024)]%
        {wikillava}
\bibfield{author}{\bibinfo{person}{Davide Caffagni}, \bibinfo{person}{Federico Cocchi}, \bibinfo{person}{Nicholas Moratelli}, \bibinfo{person}{Sara Sarto}, \bibinfo{person}{Marcella Cornia}, \bibinfo{person}{Lorenzo Baraldi}, {and} \bibinfo{person}{Rita Cucchiara}.} \bibinfo{year}{2024}\natexlab{}.
\newblock \showarticletitle{Wiki-LLaVA: Hierarchical Retrieval-Augmented Generation for Multimodal LLMs}. In \bibinfo{booktitle}{\emph{Proceedings of the IEEE/CVF Conference on Computer Vision and Pattern Recognition}}. \bibinfo{pages}{1818--1826}.
\newblock


\bibitem[Chen et~al\mbox{.}(2022)]%
        {chen2022murag}
\bibfield{author}{\bibinfo{person}{Wenhu Chen}, \bibinfo{person}{Hexiang Hu}, \bibinfo{person}{Xi Chen}, \bibinfo{person}{Pat Verga}, {and} \bibinfo{person}{William~W Cohen}.} \bibinfo{year}{2022}\natexlab{}.
\newblock \showarticletitle{Murag: Multimodal retrieval-augmented generator for open question answering over images and text}.
\newblock \bibinfo{journal}{\emph{arXiv preprint arXiv:2210.02928}} (\bibinfo{year}{2022}).
\newblock


\bibitem[Chen et~al\mbox{.}(2024)]%
        {chen2024not}
\bibfield{author}{\bibinfo{person}{Xingyu Chen}, \bibinfo{person}{Jiahao Xu}, \bibinfo{person}{Tian Liang}, \bibinfo{person}{Zhiwei He}, \bibinfo{person}{Jianhui Pang}, \bibinfo{person}{Dian Yu}, \bibinfo{person}{Linfeng Song}, \bibinfo{person}{Qiuzhi Liu}, \bibinfo{person}{Mengfei Zhou}, \bibinfo{person}{Zhuosheng Zhang}, {et~al\mbox{.}}} \bibinfo{year}{2024}\natexlab{}.
\newblock \showarticletitle{Do NOT Think That Much for 2+ 3=? On the Overthinking of o1-Like LLMs}.
\newblock \bibinfo{journal}{\emph{arXiv preprint arXiv:2412.21187}} (\bibinfo{year}{2024}).
\newblock


\bibitem[Dubey et~al\mbox{.}(2024)]%
        {dubey2024llama}
\bibfield{author}{\bibinfo{person}{Abhimanyu Dubey}, \bibinfo{person}{Abhinav Jauhri}, \bibinfo{person}{Abhinav Pandey}, \bibinfo{person}{Abhishek Kadian}, \bibinfo{person}{Ahmad Al-Dahle}, \bibinfo{person}{Aiesha Letman}, \bibinfo{person}{Akhil Mathur}, \bibinfo{person}{Alan Schelten}, \bibinfo{person}{Amy Yang}, \bibinfo{person}{Angela Fan}, {et~al\mbox{.}}} \bibinfo{year}{2024}\natexlab{}.
\newblock \showarticletitle{The llama 3 herd of models}.
\newblock \bibinfo{journal}{\emph{arXiv preprint arXiv:2407.21783}} (\bibinfo{year}{2024}).
\newblock


\bibitem[Hu et~al\mbox{.}(2024)]%
        {hu2024mrag}
\bibfield{author}{\bibinfo{person}{Wenbo Hu}, \bibinfo{person}{Jia-Chen Gu}, \bibinfo{person}{Zi-Yi Dou}, \bibinfo{person}{Mohsen Fayyaz}, \bibinfo{person}{Pan Lu}, \bibinfo{person}{Kai-Wei Chang}, {and} \bibinfo{person}{Nanyun Peng}.} \bibinfo{year}{2024}\natexlab{}.
\newblock \showarticletitle{MRAG-Bench: Vision-Centric Evaluation for Retrieval-Augmented Multimodal Models}.
\newblock \bibinfo{journal}{\emph{arXiv preprint arXiv:2410.08182}} (\bibinfo{year}{2024}).
\newblock


\bibitem[Hurst et~al\mbox{.}(2024)]%
        {hurst2024gpt}
\bibfield{author}{\bibinfo{person}{Aaron Hurst}, \bibinfo{person}{Adam Lerer}, \bibinfo{person}{Adam~P Goucher}, \bibinfo{person}{Adam Perelman}, \bibinfo{person}{Aditya Ramesh}, \bibinfo{person}{Aidan Clark}, \bibinfo{person}{AJ Ostrow}, \bibinfo{person}{Akila Welihinda}, \bibinfo{person}{Alan Hayes}, \bibinfo{person}{Alec Radford}, {et~al\mbox{.}}} \bibinfo{year}{2024}\natexlab{}.
\newblock \showarticletitle{Gpt-4o system card}.
\newblock \bibinfo{journal}{\emph{arXiv preprint arXiv:2410.21276}} (\bibinfo{year}{2024}).
\newblock


\bibitem[Izacard and Grave(2020)]%
        {rag1}
\bibfield{author}{\bibinfo{person}{Gautier Izacard} {and} \bibinfo{person}{Edouard Grave}.} \bibinfo{year}{2020}\natexlab{}.
\newblock \showarticletitle{Leveraging passage retrieval with generative models for open domain question answering}.
\newblock \bibinfo{journal}{\emph{arXiv preprint arXiv:2007.01282}} (\bibinfo{year}{2020}).
\newblock


\bibitem[Jiang et~al\mbox{.}(2024a)]%
        {jiang2024mixtral}
\bibfield{author}{\bibinfo{person}{Albert~Q Jiang}, \bibinfo{person}{Alexandre Sablayrolles}, \bibinfo{person}{Antoine Roux}, \bibinfo{person}{Arthur Mensch}, \bibinfo{person}{Blanche Savary}, \bibinfo{person}{Chris Bamford}, \bibinfo{person}{Devendra~Singh Chaplot}, \bibinfo{person}{Diego de~las Casas}, \bibinfo{person}{Emma~Bou Hanna}, \bibinfo{person}{Florian Bressand}, {et~al\mbox{.}}} \bibinfo{year}{2024}\natexlab{a}.
\newblock \showarticletitle{Mixtral of experts}.
\newblock \bibinfo{journal}{\emph{arXiv preprint arXiv:2401.04088}} (\bibinfo{year}{2024}).
\newblock


\bibitem[Jiang et~al\mbox{.}(2024b)]%
        {jiang2024mmsearch}
\bibfield{author}{\bibinfo{person}{Dongzhi Jiang}, \bibinfo{person}{Renrui Zhang}, \bibinfo{person}{Ziyu Guo}, \bibinfo{person}{Yanmin Wu}, \bibinfo{person}{Jiayi Lei}, \bibinfo{person}{Pengshuo Qiu}, \bibinfo{person}{Pan Lu}, \bibinfo{person}{Zehui Chen}, \bibinfo{person}{Guanglu Song}, \bibinfo{person}{Peng Gao}, {et~al\mbox{.}}} \bibinfo{year}{2024}\natexlab{b}.
\newblock \showarticletitle{Mmsearch: Benchmarking the potential of large models as multi-modal search engines}.
\newblock \bibinfo{journal}{\emph{arXiv preprint arXiv:2409.12959}} (\bibinfo{year}{2024}).
\newblock


\bibitem[Kil et~al\mbox{.}(2024)]%
        {kil2024ii}
\bibfield{author}{\bibinfo{person}{Jihyung Kil}, \bibinfo{person}{Farideh Tavazoee}, \bibinfo{person}{Dongyeop Kang}, {and} \bibinfo{person}{Joo-Kyung Kim}.} \bibinfo{year}{2024}\natexlab{}.
\newblock \showarticletitle{II-MMR: Identifying and improving multi-modal multi-hop reasoning in visual question answering}.
\newblock \bibinfo{journal}{\emph{arXiv preprint arXiv:2402.11058}} (\bibinfo{year}{2024}).
\newblock


\bibitem[Lee et~al\mbox{.}(2020)]%
        {lee2020learning}
\bibfield{author}{\bibinfo{person}{Jinhyuk Lee}, \bibinfo{person}{Mujeen Sung}, \bibinfo{person}{Jaewoo Kang}, {and} \bibinfo{person}{Danqi Chen}.} \bibinfo{year}{2020}\natexlab{}.
\newblock \showarticletitle{Learning dense representations of phrases at scale}.
\newblock \bibinfo{journal}{\emph{arXiv preprint arXiv:2012.12624}} (\bibinfo{year}{2020}).
\newblock


\bibitem[Lewis et~al\mbox{.}(2020)]%
        {rag2}
\bibfield{author}{\bibinfo{person}{Patrick Lewis}, \bibinfo{person}{Ethan Perez}, \bibinfo{person}{Aleksandra Piktus}, \bibinfo{person}{Fabio Petroni}, \bibinfo{person}{Vladimir Karpukhin}, \bibinfo{person}{Naman Goyal}, \bibinfo{person}{Heinrich K{\"u}ttler}, \bibinfo{person}{Mike Lewis}, \bibinfo{person}{Wen-tau Yih}, \bibinfo{person}{Tim Rockt{\"a}schel}, {et~al\mbox{.}}} \bibinfo{year}{2020}\natexlab{}.
\newblock \showarticletitle{Retrieval-augmented generation for knowledge-intensive nlp tasks}.
\newblock \bibinfo{journal}{\emph{Advances in Neural Information Processing Systems}}  \bibinfo{volume}{33} (\bibinfo{year}{2020}), \bibinfo{pages}{9459--9474}.
\newblock


\bibitem[Li et~al\mbox{.}(2022)]%
        {li_ragsurvey}
\bibfield{author}{\bibinfo{person}{Huayang Li}, \bibinfo{person}{Yixuan Su}, \bibinfo{person}{Deng Cai}, \bibinfo{person}{Yan Wang}, {and} \bibinfo{person}{Lemao Liu}.} \bibinfo{year}{2022}\natexlab{}.
\newblock \showarticletitle{A survey on retrieval-augmented text generation}.
\newblock \bibinfo{journal}{\emph{arXiv preprint arXiv:2202.01110}} (\bibinfo{year}{2022}).
\newblock


\bibitem[Li et~al\mbox{.}(2025)]%
        {li2025websailor}
\bibfield{author}{\bibinfo{person}{Kuan Li}, \bibinfo{person}{Zhongwang Zhang}, \bibinfo{person}{Huifeng Yin}, \bibinfo{person}{Liwen Zhang}, \bibinfo{person}{Litu Ou}, \bibinfo{person}{Jialong Wu}, \bibinfo{person}{Wenbiao Yin}, \bibinfo{person}{Baixuan Li}, \bibinfo{person}{Zhengwei Tao}, \bibinfo{person}{Xinyu Wang}, {et~al\mbox{.}}} \bibinfo{year}{2025}\natexlab{}.
\newblock \showarticletitle{WebSailor: Navigating Super-human Reasoning for Web Agent}.
\newblock \bibinfo{journal}{\emph{arXiv preprint arXiv:2507.02592}} (\bibinfo{year}{2025}).
\newblock


\bibitem[Lin(2004)]%
        {lin2004rouge}
\bibfield{author}{\bibinfo{person}{Chin-Yew Lin}.} \bibinfo{year}{2004}\natexlab{}.
\newblock \showarticletitle{Rouge: A package for automatic evaluation of summaries}. In \bibinfo{booktitle}{\emph{Text summarization branches out}}. \bibinfo{pages}{74--81}.
\newblock


\bibitem[Lin and Byrne(2022)]%
        {rqvqa}
\bibfield{author}{\bibinfo{person}{Weizhe Lin} {and} \bibinfo{person}{Bill Byrne}.} \bibinfo{year}{2022}\natexlab{}.
\newblock \showarticletitle{Retrieval augmented visual question answering with outside knowledge}.
\newblock \bibinfo{journal}{\emph{arXiv preprint arXiv:2210.03809}} (\bibinfo{year}{2022}).
\newblock


\bibitem[Ma et~al\mbox{.}(2023)]%
        {query_rewrite}
\bibfield{author}{\bibinfo{person}{Xinbei Ma}, \bibinfo{person}{Yeyun Gong}, \bibinfo{person}{Pengcheng He}, \bibinfo{person}{Hai Zhao}, {and} \bibinfo{person}{Nan Duan}.} \bibinfo{year}{2023}\natexlab{}.
\newblock \showarticletitle{Query rewriting for retrieval-augmented large language models}.
\newblock \bibinfo{journal}{\emph{arXiv preprint arXiv:2305.14283}} (\bibinfo{year}{2023}).
\newblock


\bibitem[Ma et~al\mbox{.}(2024)]%
        {m2rag}
\bibfield{author}{\bibinfo{person}{Zi-Ao Ma}, \bibinfo{person}{Tian Lan}, \bibinfo{person}{Rong-Cheng Tu}, \bibinfo{person}{Yong Hu}, \bibinfo{person}{Heyan Huang}, {and} \bibinfo{person}{Xian-Ling Mao}.} \bibinfo{year}{2024}\natexlab{}.
\newblock \showarticletitle{Multi-modal Retrieval Augmented Multi-modal Generation: A Benchmark, Evaluate Metrics and Strong Baselines}.
\newblock \bibinfo{journal}{\emph{arXiv preprint arXiv:2411.16365}} (\bibinfo{year}{2024}).
\newblock


\bibitem[Papineni et~al\mbox{.}(2002)]%
        {papineni2002bleu}
\bibfield{author}{\bibinfo{person}{Kishore Papineni}, \bibinfo{person}{Salim Roukos}, \bibinfo{person}{Todd Ward}, {and} \bibinfo{person}{Wei-Jing Zhu}.} \bibinfo{year}{2002}\natexlab{}.
\newblock \showarticletitle{Bleu: a method for automatic evaluation of machine translation}. In \bibinfo{booktitle}{\emph{Proceedings of the 40th annual meeting of the Association for Computational Linguistics}}. \bibinfo{pages}{311--318}.
\newblock


\bibitem[Riedler and Langer(2024)]%
        {riedler2024beyond}
\bibfield{author}{\bibinfo{person}{Monica Riedler} {and} \bibinfo{person}{Stefan Langer}.} \bibinfo{year}{2024}\natexlab{}.
\newblock \showarticletitle{Beyond Text: Optimizing RAG with Multimodal Inputs for Industrial Applications}.
\newblock \bibinfo{journal}{\emph{arXiv preprint arXiv:2410.21943}} (\bibinfo{year}{2024}).
\newblock


\bibitem[Ru et~al\mbox{.}(2024)]%
        {ru2024ragchecker}
\bibfield{author}{\bibinfo{person}{Dongyu Ru}, \bibinfo{person}{Lin Qiu}, \bibinfo{person}{Xiangkun Hu}, \bibinfo{person}{Tianhang Zhang}, \bibinfo{person}{Peng Shi}, \bibinfo{person}{Shuaichen Chang}, \bibinfo{person}{Cheng Jiayang}, \bibinfo{person}{Cunxiang Wang}, \bibinfo{person}{Shichao Sun}, \bibinfo{person}{Huanyu Li}, {et~al\mbox{.}}} \bibinfo{year}{2024}\natexlab{}.
\newblock \showarticletitle{Ragchecker: A fine-grained framework for diagnosing retrieval-augmented generation}.
\newblock \bibinfo{journal}{\emph{arXiv preprint arXiv:2408.08067}} (\bibinfo{year}{2024}).
\newblock


\bibitem[Shi et~al\mbox{.}(2023)]%
        {shi2023large}
\bibfield{author}{\bibinfo{person}{Freda Shi}, \bibinfo{person}{Xinyun Chen}, \bibinfo{person}{Kanishka Misra}, \bibinfo{person}{Nathan Scales}, \bibinfo{person}{David Dohan}, \bibinfo{person}{Ed~H Chi}, \bibinfo{person}{Nathanael Sch{\"a}rli}, {and} \bibinfo{person}{Denny Zhou}.} \bibinfo{year}{2023}\natexlab{}.
\newblock \showarticletitle{Large language models can be easily distracted by irrelevant context}. In \bibinfo{booktitle}{\emph{International Conference on Machine Learning}}. PMLR, \bibinfo{pages}{31210--31227}.
\newblock


\bibitem[Tiong et~al\mbox{.}(2022)]%
        {plugandplay}
\bibfield{author}{\bibinfo{person}{Anthony Meng~Huat Tiong}, \bibinfo{person}{Junnan Li}, \bibinfo{person}{Boyang Li}, \bibinfo{person}{Silvio Savarese}, {and} \bibinfo{person}{Steven~CH Hoi}.} \bibinfo{year}{2022}\natexlab{}.
\newblock \showarticletitle{Plug-and-play vqa: Zero-shot vqa by conjoining large pretrained models with zero training}.
\newblock \bibinfo{journal}{\emph{arXiv preprint arXiv:2210.08773}} (\bibinfo{year}{2022}).
\newblock


\bibitem[Wang et~al\mbox{.}(2023)]%
        {query2doc}
\bibfield{author}{\bibinfo{person}{Liang Wang}, \bibinfo{person}{Nan Yang}, {and} \bibinfo{person}{Furu Wei}.} \bibinfo{year}{2023}\natexlab{}.
\newblock \showarticletitle{Query2doc: Query expansion with large language models}.
\newblock \bibinfo{journal}{\emph{arXiv preprint arXiv:2303.07678}} (\bibinfo{year}{2023}).
\newblock


\bibitem[Wang et~al\mbox{.}(2024)]%
        {Qwen2VL}
\bibfield{author}{\bibinfo{person}{Peng Wang}, \bibinfo{person}{Shuai Bai}, \bibinfo{person}{Sinan Tan}, \bibinfo{person}{Shijie Wang}, \bibinfo{person}{Zhihao Fan}, \bibinfo{person}{Jinze Bai}, \bibinfo{person}{Keqin Chen}, \bibinfo{person}{Xuejing Liu}, \bibinfo{person}{Jialin Wang}, \bibinfo{person}{Wenbin Ge}, \bibinfo{person}{Yang Fan}, \bibinfo{person}{Kai Dang}, \bibinfo{person}{Mengfei Du}, \bibinfo{person}{Xuancheng Ren}, \bibinfo{person}{Rui Men}, \bibinfo{person}{Dayiheng Liu}, \bibinfo{person}{Chang Zhou}, \bibinfo{person}{Jingren Zhou}, {and} \bibinfo{person}{Junyang Lin}.} \bibinfo{year}{2024}\natexlab{}.
\newblock \showarticletitle{Qwen2-VL: Enhancing Vision-Language Model's Perception of the World at Any Resolution}.
\newblock \bibinfo{journal}{\emph{arXiv preprint arXiv:2409.12191}} (\bibinfo{year}{2024}).
\newblock


\bibitem[Wu et~al\mbox{.}(2025)]%
        {wu2025webdancer}
\bibfield{author}{\bibinfo{person}{Jialong Wu}, \bibinfo{person}{Baixuan Li}, \bibinfo{person}{Runnan Fang}, \bibinfo{person}{Wenbiao Yin}, \bibinfo{person}{Liwen Zhang}, \bibinfo{person}{Zhengwei Tao}, \bibinfo{person}{Dingchu Zhang}, \bibinfo{person}{Zekun Xi}, \bibinfo{person}{Gang Fu}, \bibinfo{person}{Yong Jiang}, {et~al\mbox{.}}} \bibinfo{year}{2025}\natexlab{}.
\newblock \showarticletitle{WebDancer: Towards Autonomous Information Seeking Agency}.
\newblock \bibinfo{journal}{\emph{arXiv preprint arXiv:2505.22648}} (\bibinfo{year}{2025}).
\newblock


\bibitem[Wu et~al\mbox{.}(2017)]%
        {vqa_survey}
\bibfield{author}{\bibinfo{person}{Qi Wu}, \bibinfo{person}{Damien Teney}, \bibinfo{person}{Peng Wang}, \bibinfo{person}{Chunhua Shen}, \bibinfo{person}{Anthony Dick}, {and} \bibinfo{person}{Anton Van Den~Hengel}.} \bibinfo{year}{2017}\natexlab{}.
\newblock \showarticletitle{Visual question answering: A survey of methods and datasets}.
\newblock \bibinfo{journal}{\emph{Computer Vision and Image Understanding}}  \bibinfo{volume}{163} (\bibinfo{year}{2017}), \bibinfo{pages}{21--40}.
\newblock


\bibitem[Yuan et~al\mbox{.}(2023)]%
        {yuan2023ramm}
\bibfield{author}{\bibinfo{person}{Zheng Yuan}, \bibinfo{person}{Qiao Jin}, \bibinfo{person}{Chuanqi Tan}, \bibinfo{person}{Zhengyun Zhao}, \bibinfo{person}{Hongyi Yuan}, \bibinfo{person}{Fei Huang}, {and} \bibinfo{person}{Songfang Huang}.} \bibinfo{year}{2023}\natexlab{}.
\newblock \showarticletitle{Ramm: Retrieval-augmented biomedical visual question answering with multi-modal pre-training}. In \bibinfo{booktitle}{\emph{Proceedings of the 31st ACM International Conference on Multimedia}}. \bibinfo{pages}{547--556}.
\newblock


\bibitem[Zhang et~al\mbox{.}(2024a)]%
        {zhang2024path}
\bibfield{author}{\bibinfo{person}{Ge Zhang}, \bibinfo{person}{Mohammad~Ali Alomrani}, \bibinfo{person}{Hongjian Gu}, \bibinfo{person}{Jiaming Zhou}, \bibinfo{person}{Yaochen Hu}, \bibinfo{person}{Bin Wang}, \bibinfo{person}{Qun Liu}, \bibinfo{person}{Mark Coates}, \bibinfo{person}{Yingxue Zhang}, {and} \bibinfo{person}{Jianye Hao}.} \bibinfo{year}{2024}\natexlab{a}.
\newblock \showarticletitle{Path-of-Thoughts: Extracting and Following Paths for Robust Relational Reasoning with Large Language Models}.
\newblock \bibinfo{journal}{\emph{arXiv preprint arXiv:2412.17963}} (\bibinfo{year}{2024}).
\newblock


\bibitem[Zhang et~al\mbox{.}(2024b)]%
        {zhang2024mr2ag}
\bibfield{author}{\bibinfo{person}{Tao Zhang}, \bibinfo{person}{Ziqi Zhang}, \bibinfo{person}{Zongyang Ma}, \bibinfo{person}{Yuxin Chen}, \bibinfo{person}{Zhongang Qi}, \bibinfo{person}{Chunfeng Yuan}, \bibinfo{person}{Bing Li}, \bibinfo{person}{Junfu Pu}, \bibinfo{person}{Yuxuan Zhao}, \bibinfo{person}{Zehua Xie}, {et~al\mbox{.}}} \bibinfo{year}{2024}\natexlab{b}.
\newblock \showarticletitle{mR$^{2}$ AG: Multimodal Retrieval-Reflection-Augmented Generation for Knowledge-Based VQA}.
\newblock \bibinfo{journal}{\emph{arXiv preprint arXiv:2411.15041}} (\bibinfo{year}{2024}).
\newblock


\bibitem[Zhang et~al\mbox{.}(2023)]%
        {zhang2023sentiment}
\bibfield{author}{\bibinfo{person}{Wenxuan Zhang}, \bibinfo{person}{Yue Deng}, \bibinfo{person}{Bing Liu}, \bibinfo{person}{Sinno~Jialin Pan}, {and} \bibinfo{person}{Lidong Bing}.} \bibinfo{year}{2023}\natexlab{}.
\newblock \showarticletitle{Sentiment analysis in the era of large language models: A reality check}.
\newblock \bibinfo{journal}{\emph{arXiv preprint arXiv:2305.15005}} (\bibinfo{year}{2023}).
\newblock


\bibitem[Zhao et~al\mbox{.}(2023a)]%
        {mrag_survey}
\bibfield{author}{\bibinfo{person}{Ruochen Zhao}, \bibinfo{person}{Hailin Chen}, \bibinfo{person}{Weishi Wang}, \bibinfo{person}{Fangkai Jiao}, \bibinfo{person}{Xuan~Long Do}, \bibinfo{person}{Chengwei Qin}, \bibinfo{person}{Bosheng Ding}, \bibinfo{person}{Xiaobao Guo}, \bibinfo{person}{Minzhi Li}, \bibinfo{person}{Xingxuan Li}, {et~al\mbox{.}}} \bibinfo{year}{2023}\natexlab{a}.
\newblock \showarticletitle{Retrieving multimodal information for augmented generation: A survey}.
\newblock \bibinfo{journal}{\emph{arXiv preprint arXiv:2303.10868}} (\bibinfo{year}{2023}).
\newblock


\bibitem[Zhao et~al\mbox{.}(2023b)]%
        {zhao2023retrieving}
\bibfield{author}{\bibinfo{person}{Ruochen Zhao}, \bibinfo{person}{Hailin Chen}, \bibinfo{person}{Weishi Wang}, \bibinfo{person}{Fangkai Jiao}, \bibinfo{person}{Xuan~Long Do}, \bibinfo{person}{Chengwei Qin}, \bibinfo{person}{Bosheng Ding}, \bibinfo{person}{Xiaobao Guo}, \bibinfo{person}{Minzhi Li}, \bibinfo{person}{Xingxuan Li}, {et~al\mbox{.}}} \bibinfo{year}{2023}\natexlab{b}.
\newblock \showarticletitle{Retrieving multimodal information for augmented generation: A survey}.
\newblock \bibinfo{journal}{\emph{arXiv preprint arXiv:2303.10868}} (\bibinfo{year}{2023}).
\newblock


\bibitem[Zheng et~al\mbox{.}(2024)]%
        {zheng2024llamafactory}
\bibfield{author}{\bibinfo{person}{Yaowei Zheng}, \bibinfo{person}{Richong Zhang}, \bibinfo{person}{Junhao Zhang}, \bibinfo{person}{Yanhan Ye}, \bibinfo{person}{Zheyan Luo}, \bibinfo{person}{Zhangchi Feng}, {and} \bibinfo{person}{Yongqiang Ma}.} \bibinfo{year}{2024}\natexlab{}.
\newblock \showarticletitle{LlamaFactory: Unified Efficient Fine-Tuning of 100+ Language Models}. In \bibinfo{booktitle}{\emph{Proceedings of the 62nd Annual Meeting of the Association for Computational Linguistics (Volume 3: System Demonstrations)}}. \bibinfo{publisher}{Association for Computational Linguistics}, \bibinfo{address}{Bangkok, Thailand}.
\newblock


\bibitem[Zhu et~al\mbox{.}(2024)]%
        {zhu2024murar}
\bibfield{author}{\bibinfo{person}{Zhengyuan Zhu}, \bibinfo{person}{Daniel Lee}, \bibinfo{person}{Hong Zhang}, \bibinfo{person}{Sai~Sree Harsha}, \bibinfo{person}{Loic Feujio}, \bibinfo{person}{Akash Maharaj}, {and} \bibinfo{person}{Yunyao Li}.} \bibinfo{year}{2024}\natexlab{}.
\newblock \showarticletitle{Murar: A simple and effective multimodal retrieval and answer refinement framework for multimodal question answering}.
\newblock \bibinfo{journal}{\emph{arXiv preprint arXiv:2408.08521}} (\bibinfo{year}{2024}).
\newblock


\end{thebibliography}

\end{document}